\documentclass[aps,pra,twocolumn,superscriptaddress,showpacs,floatfix]{revtex4-1}

\usepackage{amsmath,amssymb}
\usepackage{graphicx,float}
\usepackage{bm}
\usepackage{mathrsfs}
\usepackage{rotating}
\usepackage[amssymb]{SIunits}
\usepackage{inputenc}
\usepackage{multirow}
\usepackage{tabularx}
\usepackage{color}

\allowdisplaybreaks[2]

\begin{document}

\title{Detection of a weak magnetic field via cavity enhanced Faraday rotation}

 \author{Keyu \surname{Xia}}  %
  \email{keyu.xia@mq.edu.au}
 \affiliation{ARC Centre for Engineered Quantum Systems, Department of Physics and Astronomy, Macquarie University, NSW 2109, Australia}

 \author{Nan \surname{Zhao}}
   \email{nzhao@csrc.ac.cn }
 \affiliation{Beijing Computational Science Research Center, Beijing 100084, China}

 \author{Jason \surname{Twamley}}
  \email{jason.twamley@mq.edu.au}
 \affiliation{ARC Centre for Engineered Quantum Systems, Department of Physics and Astronomy, Macquarie University, NSW 2109, Australia}

\begin{abstract}
We study the sensitive detection of a weak static magnetic field via Faraday rotation induced by an ensemble of spins in a bimodal degenerate microwave cavity. We determine the limit of the resolution for the sensitivity of the magnetometry achieved using either single-photon or multiphoton inputs. For the case of a microwave cavity containing an ensemble of Nitrogen-vacancy defects in diamond, we obtain a magnetometry sensitivity exceeding $0.5~\text{\nano\tesla}/\sqrt{\text{\hertz}}$, utilizing a single photon probe field, while for a multiphoton input we achieve a sensitivity about $1 \text{\femto\tesla}/\sqrt{\text{\hertz}}$, using a coherent probe microwave field with power of $P_\text{in}=1~\text{\nano\watt}$.
\end{abstract}

% 33.57.+c Faraday effect
% 42.50.Ct Quantum description of interaction of light and matter; related experiments
% 42.50.Pq Cavity QED
% Magnetometers, 07.55.Ge, 07.55.Jg
% 07.55.Ge Magnetometers for magnetic field measurements
% Magneto-optical devices, 85.70.Sq
% Magneto-optical effects, 78.20.Ls
% 06.20.-f Metrology

\maketitle
%Introduction
\section{Introduction}
Sensing of magnetic fields with an extremely high sensitivity has attractive applications in various areas of science and technology \cite{MagnetometryRev1,MagnetometryRev2,MagnetometryRev3}. Magnetic fields can be measured using a variety of techniques including a superconducting quantum interference device (SQUID) \cite{MagnetometryRev1}, magneto-materials \cite{PRLBowen}, atoms\cite{AtomicMagnetometry1,AtomicMagnetometry2,AtomicMagnetometry3,AtomicMagnetometry4} and color defect centres in diamond \cite{DiamondMagnetometry1,DiamondMagnetometry2,DiamondMagnetometry3}. The SQUID magnetometers achieve record sensitivities \cite{MagnetometryRev1}, but require extremely low temperatures to maintain superconductivity \cite{MagnetometryRev1}. Recently, atomic magnetometers have demonstrated subfemtotesla sensitivity, approaching the record sensitivity of SQUID sensors \cite{AtomicMagnetometry2,AtomicMagnetometry3,AtomicMagnetometry4}. Diamond magnetometers exploiting nitrogen vacancy (NV) defect centres in diamond offers detection of magnetic field signals both with high spatial accuracy \cite{NanoDiamondMagnetometry1,NanoDiamondMagnetometry2,NanoDiamondMagnetometry3,MagnetometryNanZhao} as well as high field sensitivity down to sub-$\text{\pico\tesla}/\sqrt{\text{\hertz}}$ \cite{DiamondMagnetometry1,DiamondMagnetometry3}. In this work, we describe a magnetometer approaching sub-$\text{\femto\tesla}/\sqrt{\text{\hertz}}$ sensitivity with $\nano\watt$-order input power that exploits the Faraday rotation of microwave (mw) photons induced by a static magnetic field.

In the Faraday effect the polarization of electromagnetic fields traveling in magneto-optical material can be rotated by applying a static magnetic field. This Faraday rotation can also be well explained by a quantum mechanism \cite{FaradayRotation1}. Based on this quantum understanding Faraday rotation has been proposed to detect quantum fluctuations \cite{FaradayRotation2}, and induce giant phase modulation \cite{PhysRevA.70.023822}. Giant optical Faraday rotation has also been observed \cite{GiantFaraday1,GiantFaraday2}. The polarization rotation due to the Faraday effect has been well studied for classical optical magnetometers but a corresponding discussion of the sensitivity limit is so far absent in the literature.

In this paper we investigate the sensitivity limit of a microwave version of a Faraday magnetometer. In our scheme the polarization of microwave probe photons is rotated by an ensemble of spins coupled to a microwave cavity. By measuring the subsequent rotation of the photon's polarization we are able to detect a weak static magnetic field applied to the spins. Thanks to the low frequency of microwave photons and the lower vacuum energy fluctuations, as compared with the optical photons, we find that the microwave reflectance Faraday magnetometer is ultrasensitive in comparison with optical magnetometers, given the same probe field power.

\section{System and model}
Before describing the details of our system and model we first present the main idea of our scheme to measure a weak magnetic field. Our scheme involves an ensemble of spins coupled to a mw cavity. A horizontal-polarized probe mw field is incident to the cavity. This mw input field can be decomposited into the $\sigma_+$- (right circular-polarized) and $\sigma_-$-polarized (left circular-polarized) components. The mw cavity also possesses $\sigma_+$- and $\sigma_-$-polarized modes which are degenerate in frequency. The two orthogonally polarized cavity modes couple to two separate transitions of spins contained in the cavity. When a weak static magnetic field is applied to the ensemble of spins it shifts the two transition energies up or down oppositely and subsequently causes different detunings between the transitions and the cavity modes. Thus, the reflected fields suffers different phase shifts yielding a Faraday rotation angle. As a result, in the output the superposition field of the two polarization-orthogonal modes includes both horizontal and vertical polarized photons. Measuring these photons we can estimate the magnetic field with a high precision.

\begin{figure}
\centering
\setlength{\unitlength}{1cm}
\includegraphics[width=0.8\linewidth]{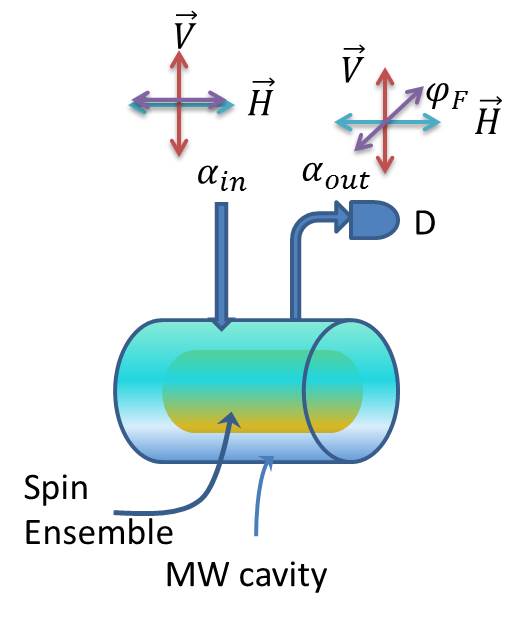}\\
\caption{(Color online) Schematics of setup for detection of a weak static magnetic field $B$. An ensemble of spins couples to a microwave (mw) cavity. The horizontal polarized mw field $\alpha_\text{in}$ inputs into the mw cavity and then is reflected off the cavity ($\alpha_{out}$) to the detector. The output field reflected off the cavity is detected by the polarization resolving photon detector D. The polarization of the output field is rotated by an angle $\varphi_F$ due to the Faraday effect. The cavity has the resonance frequency $\omega_r$ and the intrinsic loss $\kappa_\text{i})$. The input mw field couples to the cavity with an external coupling rate $\kappa_\text{ex}$.}\label{fig:setup}
\end{figure}

\begin{figure}
\centering
\setlength{\unitlength}{1cm}
\includegraphics[width=0.99\linewidth]{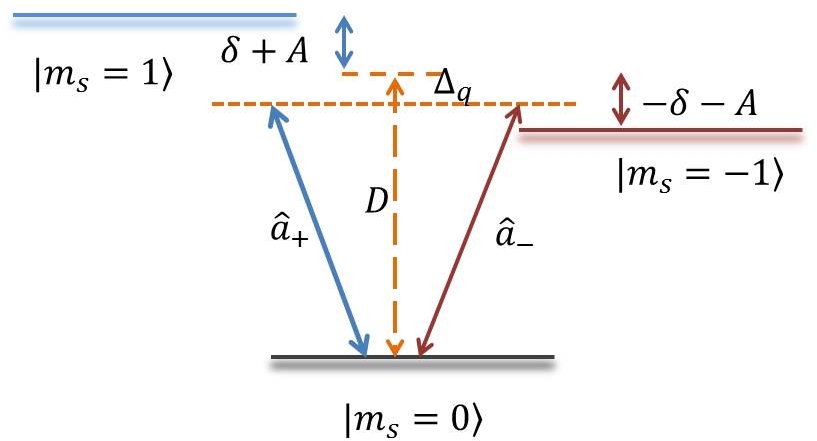}\\
\caption{(Color online) Configuration describing the interaction between the cavity mode $\hat{a}_\pm$ and an ensemble of NV$^-$ centers. Two degenerate cavity modes, the right circular-polarized mode $\hat{a}_+$ and the left circular-polarized mode $\hat{a}_-$, are detuned from the zero-strain splitting by $\Delta_q$. We shift the transition $|m_s=1\rangle \leftrightarrow |m_s=0\rangle (|m_s=-1\rangle \leftrightarrow |m_s=0\rangle)$ by a bias static magnetic field $B_0$, which creates a static frequency shift $A=\mu_B g_eB_0$. The weak static magnetic field $\delta B$ causes another small shift $\delta=\mu_B g_e \delta B$.}\label{fig:LevelScheme}
\end{figure}

We now describe in more detail our scheme for magnetic field sensing. In the scheme an ensemble of NV centers couples to the mw cavity as shown in Fig. \ref{fig:setup}. 
We assume that the mw cavity supports right and left circular-polarized ($\sigma_+$ and $\sigma_-$-polarized) cavity modes, $\{\hat{a}_{+},\hat{a}_{-}\}$, and these modes are degenerate with resonance frequency $\omega_r$, and intrinsic loss $\kappa_\text{i}$.  A horizontal ($\text{H}$)-polarized coherent mw field $\alpha_\text{in}$ with frequency $\omega_\text{in}$ enters the cavity with an external coupling rate of $\kappa_\text{ex}$ and then yields an output field $\alpha_{out}$. This input field, $\alpha_\text{in}$, can be decomposed into $\sigma_+$ and $\sigma_-$-polarized components driving the cavity modes. The $\sigma_+$ and $\sigma_-$-polarized cavity fields suffer different phase shift because they couple to different transition within the spin ensemble (Fig. \ref{fig:LevelScheme}). Due to the interaction between the spins and the mw fields in cavity, the polarization of the output field  $\alpha_\text{out}$ is rotated by the Faraday angle, $\phi_F$, and hence $\alpha_{out}$ includes a vertical ($\text{V}$)-polarized component. For a clear discussion of the relation between the input and output amplitudes and polarizations we define the annihilation and creation operators, $\hat{\varsigma}$ and $\hat{\varsigma}^\dag$, for a quantized radiation field. We use the notation $\hat{\varsigma}\vec{\chi}$ \cite{ScullyQO,MilburnQO}, to denote a $\chi-$polarized field of mode $\hat{\varsigma}$, where $\vec{\chi}$ is the unit polarization vector. The polarization vector $\vec{\chi}$ can be $\vec{\text{H}},\vec{\text{V}},\vec{\sigma_+},\vec{\sigma_-}$ or an arbitrary linear polarization $\vec{\theta} = \cos\theta \vec{\text{H}} + \sin\theta \vec{\text{V}}$ with $\theta \in (-\pi,\pi]$. Next we will discuss the magnetometer sensitivity in the case of single-photon and multiphoton measurements.

In the case of a very weak input a single-photon probe mw field incidents to the cavity and we consider all possible outputs. The output mw field is no longer perfectly horizontal polarized due to the Faraday effect and a polarization sensitive mw photon detector connected to the output port may detect a horizontal or vertical polarized photon or a ``dark count" with associated probabilities. The ``dark count" implies that the input photon is lost to the environment or is absorbed by the materials. The limit of the sensitivity can be determined by the Fisher information.
In classical measurement, we input a weak coherent probe mw field with power $P_\text{in}$ into the cavity and only measure  the intensity of the vertical-polarized output component $I_{\vec{V}}$.
 The horizontal-polarized component is filtered from the output. 

Our scheme involves the interaction of mw modes in the cavity and an ensemble of spins, taken here, for example to be NV centers. In Fig. \ref{fig:LevelScheme} we graph the ground state triplet of the NV defect in the presence of a static bias field. The mw cavity modes $\hat{a}_+$ and $\hat{a}_-$ drive the magnetic transitions $|m_s=1\rangle \leftrightarrow |m_s=0\rangle$ and $|m_s=-1\rangle \leftrightarrow |m_s=0\rangle$, respectively. In the absence of any magnetic field, the zero-strain splitting of the NV centers is assumed to be $D\approx 2.78$ \giga\hertz. Under the bias magnetic field $B_0$ and the signal field to be sensed, $\delta B$, the levels of $|m_s=1\rangle$ and $|m_s=-1\rangle$ are shifted up or down by $A+\delta$, respectively. The magnetic field is applied along the $z$-axis of the spin crystal, and results in $\delta=\mu_B g_e \delta B$ and $A=\mu_B g_e B_0$, where $\mu_B=14$ \mega\hertz \cdot \milli\tesla$^{-1}$ is the Bohr magneton and $g_e\approx 2$ the electron spin $g$-factor. The interaction Hamiltonian describing the coupling between the spins and cavity takes the form
\begin{equation}\nonumber
\begin{split}
 \hat{H}_\text{I}=&\sum_j \left(g_{j,R} \hat{a}_+^\dag |m_s=0\rangle_j \langle m_s=1| + H.c.\right) \\
&+ \sum_j \left(g_{j,L} \hat{a}_-^\dag |m_s=0\rangle_j \langle m_s=-1| + H.c. \right) \,,
\end{split}
\end{equation}
where $g_{j,R}$ ($g_{j,L}$) is the coupling rate between the cavity mode $\hat{a}_{+}$ ($\hat{a}_{-}$) and the transition $|m_s=1\rangle \leftrightarrow |m_s=0\rangle$ ($|m_s=-1\rangle \leftrightarrow |m_s=0\rangle$) of the $j$th spin. For simplicity we assume that $g_{j,R}=g_{j,L}=g$ for all spins. For an ensemble of $N$ spins we can apply the Holstein-Primakoff transformation \cite{HFTransform} to define the collective operators $\hat{c}_{+}=1/\sqrt{N} \sum_j ^N |m_s=0\rangle_j \langle m_s=1|$ and $\hat{c}_{-}=1/\sqrt{N} \sum_j ^N |m_s=0\rangle_j \langle m_s=-1|$, which allows one to consider a large number of spins as a generalized harmonic oscillator coupled to the cavity mode with a collectively enhanced rate $G=\sqrt{N}g$. The Holstein-Primakoff transformation requires that the spin ensemble is highly polarized in the $|m_s\rangle$ state which can be achieved via optical pumping. Hence, the interaction Hamiltonian now becomes 
\begin{equation}
 H_\text{I}=\left(G \hat{a}_+^\dag \hat{c}_+ + G^* \hat{a}_-^\dag \hat{c}_-  + H.c. \right) \,.
\end{equation}
The free evolution Hamiltonian of spins under the magnetic fields $B_0$ and $\delta B$ is given by
\begin{equation} \nonumber
 H_\text{spin} = \sum_j D_j S_{z,j}^2 + \sum_j (A+\delta) S_{z,j} \,,
\end{equation}
where $S_{z,j}$ is the $z$ component of the spin-1 operator, and $D_j$ the zero-strain splitting of the $j$th spin. If we neglect the inhomogeneous broadening of $D_j$ due to different local strains and assume $D_j=D$, then we have 
\begin{equation}
 H_{\text{spin}} = D (\hat{c}_{+}^\dag\hat{c}_{+} + \hat{c}_-^{\dag}\hat{c}_{-}) + (A+\delta) (\hat{c}_{+}^\dag\hat{c}_{+} - \hat{c}_-^{\dag}\hat{c}_{-})\,.
\end{equation}

We consider input probe fields $\hat{a}^\pm_\text{in}$, each with frequency $\omega_\text{in}$. 
The full Hamiltonians for $\sigma_+$ and $\sigma_-$-polarized cavity modes and spin ensemble takes the uniform form
\begin{align}\label{eq:fullH}
 H_{\text{full}}^\pm/\hbar = & \omega_{r} \hat{a}_\pm^\dag \hat{a}_\pm + D \hat{c}^\dag_\pm \hat{c}_\pm \pm (A+\delta) \hat{c}_\pm^\dag \hat{c}_\pm  \;\\ \nonumber
& + (G \hat{a}^\dag_\pm \hat{c}_\pm + G^* \hat{c}^\dag_\pm \hat{a}_\pm)  \;\\ \nonumber
 & + i\sqrt{2\kappa_{\text{ex}}} (\hat{a}^\pm_\text{in} \hat{a}_\pm^\dag -\hat{a}_\text{in}^{\pm,\dag} \hat{a}_\pm) \;,
\end{align}
where the $\sigma_+$($\sigma_-$)-polarized input uses the $+$($-$) sign, and $\gamma$ is the decoherence rate of the spin ensemble for each transition.
Using these Hamiltonians and the input-output relation \cite{InputOutputPhysRevA.30.1386,InputOutputPhysRevA.31.3761}, we can calculate the amplitude of the reflection in the frequency domain at $\omega=\omega_\text{in}$ to be %\cite{RefOptom1,RefOptom2}
\begin{equation}\label{eq:r}
 r_\pm(\omega_\text{in}) =  -1 + \frac{2\kappa_{ex}}{i\Delta_r + (\kappa_{ex} + \kappa_{i}) + \frac{G^2}{i[\Delta_q \pm (A+\delta)] + \frac{\gamma}{2}}}\,,
\end{equation}
where the detunings $\Delta_r = \omega_r - \omega_\text{in}$ and $\Delta_q = D - \omega_\text{in}$. $\omega_\text{in}$ is the carrier frequency of the input probe field of $\alpha_\text{in}$.
Equation (\ref{eq:r}) is valid for both a single-photon and classical probe field when the number of photons in the cavity is much smaller than the number of spins \cite{RefOptom1,RefOptom2,Probability}. We convert the input and output fields from the $\sigma_+$ and $\sigma_-$-polarized basis to the $\text{H}$ and $\text{V}$-polarized basis by the relations $\vec{\sigma}_+=(\vec{H}-i\vec{V})/\sqrt{2}, \vec{\sigma}_-=(\vec{H}+i\vec{V})/\sqrt{2}$ 
[or $\vec{H}=(\vec{\sigma}_+ + \vec{\sigma}_-)/\sqrt{2},~\vec{V}=i(\vec{\sigma}_+ - \vec{\sigma}_-)/\sqrt{2}$] \cite{PRelation}. Thus, the input and output in the $\text{H}$ and $\text{V}$-polarized basis is governed by a scattering matrix as
\begin{equation}\label{eq:Sr}
\begin{pmatrix}
 \hat{a}^\text{H}_\text{out}\vec{H}  \\
 \hat{a}^\text{V}_\text{out}\vec{V} 
\end{pmatrix} 
=
S_r
\begin{pmatrix}
 \hat{a}^\text{H}_\text{in}\vec{H} \\
 \hat{a}^\text{V}_\text{in}\vec{V}
\end{pmatrix} \;,
\end{equation}
with
\begin{equation}
 S_r=\begin{pmatrix}
 r_{\vec{H}\vec{H}} & ir_{\vec{H}\vec{V}} \\
 -ir_{\vec{V}\vec{H}} & r_{\vec{V}\vec{V}}
\end{pmatrix}
=\begin{pmatrix}
 \frac{r_+ + r_-}{2} & i \frac{r_+ - r_-}{2} \\
 -i\frac{r_+ - r_-}{2} &  \frac{r_+ + r_-}{2} 
\end{pmatrix} \;.
\end{equation}

Specifically, when a $\text{H}$-polarized mw probe field $\hat{a}^\text{H}_\text{in} \vec{H}=\hat{a}^+_\text{in}\vec{\sigma}_+ + \hat{a}^-_\text{in}\vec{\sigma}_- = \hat{a}^\text{H}_\text{in}/\sqrt{2}(\vec{\sigma}_+ +\vec{\sigma}_-)$ is input into the cavity, the output field takes the form $\hat{a}_\text{out}\vec{\chi}_\text{out}=\frac{\hat{a}_\text{in}}{2} (r_+ + r_-)\vec{H} - i\frac{\hat{a}_\text{in}}{2} (r_+ - r_-)\vec{V}$. The corresponding output power spectrum of the vertical and horizontal polarization are given by $S_{\vec{V}}(\omega_\text{in})= R_\text{V}\langle \hat{a}^{\text{H}\dag}_\text{in}(-\omega_\text{in}) \hat{a}^\text{H}_\text{in}(\omega_\text{in})\rangle$ and $S_{\vec{H}}(\omega_\text{in})=R_\text{H}\langle \hat{a}^{\text{H}\dag}_\text{in}(-\omega_\text{in}) \hat{a}^\text{H}_\text{in}(\omega_\text{in})\rangle $ \cite{PSD3}, respectively, where $R_{\text{V}}=|r_{\vec{V}\vec{H}}|^2 (R_\text{H}=|r_{\vec{H}\vec{H}}|^2)$.
We define the reflection $r_{\pm}=|r_{\pm}|e^{i\varphi_{\pm}}$, $\bar{r}=\frac{|r_+|+|r_-|}{2}$, and $\delta{r}=\frac{|r_+|-|r_-|}{2}$. The polarization of the output field is rotated by Faraday angle $\varphi_F=\frac{\varphi_+ - \varphi_-}{2}$ with respect to the $\text{H}$-polarized input field.
 For a $\text{V}$-polarized input field $\hat{a}^\text{V}_\text{in}\vec{V} = i \hat{a}^\text{V}_\text{in}/\sqrt{2} (\vec{\sigma}_+ - \vec{\sigma}_-)$, the output field is $i\frac{\hat{a}^\text{V}_\text{in}}{2} (r_+ - r_-)\vec{H}+ \frac{\hat{a}^\text{V}_\text{in}}{2} (r_+ + r_-)\vec{V}$.

\section{Limit of magnetometry sensitivity}
\subsection{Single-photon input}
If we assume that our detector has a unit quantum efficiency $\eta=1$ and input a single photon into the setup, the probability to detect an output photon polarized along $\vec{V}$ or $\vec{H}$ can be determined as $P_{\vec{V}}=S_{\vec{V}}/S_\text{in}$ or $P_{\vec{H}}=S_{\vec{H}}/S_\text{in}$ with $S_\text{in}(\omega)=\langle \hat{a}^\dag_\text{in}(-\omega_\text{in}) \hat{a}_\text{in}(\omega_\text{in})\rangle$ at $\omega=\omega_\text{in}$. It is also possible that no photon clicks the detector due to the ``dark count" indicating the loss of the photon before clicking the detector. This latter probability due to the loss of photon from cavities is $P_{\O{}}=1-P_{\vec{V}}-P_{\vec{H}}$. The probabilities for three possible outputs are
\begin{equation} \label{eq:Prob}
 \begin{split}
 P_{\vec{V}} &= \bar{r}^2 \sin^2(\varphi_F) + \delta r^2 \cos^2(\varphi_F)=|r_{\vec{V}\vec{H}}|^2 \;,\\
 P_{\vec{H}} &= \bar{r}^2 \cos^2(\varphi_F) + \delta r^2 \sin^2(\varphi_F)=|r_{\vec{H}\vec{H}}|^2 \;,\\
 P_{\O{}} &=1-(\bar{r}^2 + \delta r^2) \;.
\end{split}
\end{equation}

We are interested in the sensitivity of the system to perform a measurement of the static signal field $\delta B$, $\frac{\partial P(\xi|\delta B)}{\partial \delta B}$, which means how fast the probabilities to detect a photon in state $\xi=\{\vec{V},\vec{H},\O{}\}$ change for a certain signal $\delta B$. We first examine the limit of the magnetometry sensitivity using a single-photon input mw pulse. Consider a single-photon probe prepared in an initial quantum state $\rho(0)$ that this evolves to a state $\rho(\tau_m)$ when exposed to the signal $\delta=\mu_B g_e \delta B$ after a measurement time $\tau_m$. This generates three possible outcomes with probabilities $P_{\xi}$ with $\xi\in\{\vec{H},\vec{V},\O{}\}$. To evaluate the performance of sensitivity of our setup, we rescale all parameters by $\kappa_{i}$. Generally, the maximum amount of information about $\delta$ that can be extracted from the polarization dependent measurement is given by the Fisher information \cite{FisherInf1,FisherInf2}
\begin{equation}\label{eq:FishInf}
\begin{split}
 F_I(\delta B) & = \left(\frac{\mu_B g_e}{\kappa_{i}}\right)^2\sum_{\xi=\vec{V},\vec{H},\O{}} \frac{1}{P(\xi|\delta/\kappa_{i})} \left( \frac{\partial P(\xi|\delta/\kappa_{i})}{ \partial \delta /\kappa_{i} }\right)^2 \;\\
 & = \left(\mu_B g_e\right)^2\sum_{\xi=\vec{V},\vec{H},\O{}} \frac{1}{P(\xi|\delta)} \left( \frac{\partial P(\xi|\delta)}{ \partial \delta }\right)^2 \;.
\end{split}
\end{equation}
This leads to the Cram\'{e}r-Rao bound \cite{CRBound1,FisherInf2}
\begin{equation}\label{eq:CRBoundB}
 \Delta B \geq \frac{1}{\sqrt{\nu_m F_I(\delta B)}} \;,
\end{equation}
where $\nu_m$ is the number of times the measurement is repeated. If the total measurement time is $\tau_\text{total}$ and each measurement takes a time $\tau_m$, then we have the following sensitivity for a single-photon input
\begin{equation}\label{eq:SPLimit}
 \Delta B_\text{SP} \sqrt{\tau_\text{total}} \geq \frac{\sqrt{\tau_m}}{\sqrt{F_I(\delta B)}} \;.
\end{equation}
 In practice the time $\tau_m$ is determined by the full width at half maximum (FWHM) of the Fisher information when evaluated as a function of the signal $\delta B$ (See Fig. \ref{fig:FIdelta} below). 

\subsection{Multiphoton input}
Generally, taken estimating the value of a parameter $\varphi$ contained in a measurement operator $\hat{M}$, the root mean square (RMS), $\Delta\varphi$, of variance as an imprecision in the estimation can be obtained from  \cite{VarM1,VarM2,VarM3}
\begin{equation} \label{eq:VarLim}
 \Delta\varphi^2 = \frac{\Delta \hat{M}^2}{|\partial \langle \hat{M}\rangle /\partial \varphi|^2} \;,
\end{equation}
where the variance $\Delta \hat{M}^2 \equiv \langle \hat{M}^2 \rangle - \langle \hat{M} \rangle^2$, with the expectation values taken as the appropriate input state. We only detect the $\text{V}$-polarized output photons so we have $\hat{M}= \hat{a}^{\dag\text{V}}_\text{out}\hat{a}^\text{V}_\text{out}$. 

We now estimate the limit of the measurement sensitivity for the multiphoton input where the probe field is a weak coherent mw pulse with an input power $P_\text{in}$ corresponding to a mean photon number $\bar{n}_\text{in}$, the quantum expectation of the operator $\hat{n}_\text{in} = \hat{a}^{\text{H}\dag}_\text{in} \hat{a}^\text{H}_\text{in}$. We have $\langle \hat{a}^{^\text{H}\dag}_\text{in} \hat{a}^\text{H}_\text{in} \rangle = 2\kappa_\text{ex}\bar{n}_\text{in} = \frac{\tau_m}{\tau_\text{ex}} \frac{P_\text{in}}{\hbar\omega_r}$ \cite{PhysRevLett.99.093901,Science319p1062} ($\bar{n}_\text{in} =\frac{P_\text{in}\tau_m}{\hbar\omega_r}$ \cite{PhysRevA.70.023822}), 
where $1/\tau_{ex}=2\kappa_\text{ex}$ is the photon decay rate into the associated outgoing modes \cite{PhysRevLett.99.093901}. Here we replace $\tau_\text{ex}$ with the duration $\tau_m$ of the probe pulse when calculating $n_\text{in}$ \cite{PhysRevLett.99.093901} because the duration of probe pulse is limited by the bandwidth determined by Fisher information now. The spectrum of the input probe mw field $S_\text{in}(\omega_\text{in})=\langle \hat{a}^{\text{H}\dag}_\text{in}(-\omega_\text{in})\hat{a}^\text{H}_\text{in}(\omega_\text{in})\rangle= 2\kappa_\text{ex} \bar{n}_\text{in}$ \cite{PSD3,PSD4}. We now focus on the vertical polarized output. 
 
%% explain noise sources
In our setup, the input and output ports, and the cavity support both $\text{H}$- and $\text{V}$-polarized fields. As a result, both the $\text{H}$- and $\text{V}$-polarized noise can enter the input-output port and the cavity through the external coupling channel or the intrinsic lossy channel, and then are reflected to the detector. The operators denoting the noise entering the cavities are $\hat{\xi}^\text{H}_\text{E}\vec{H}=\hat{\xi}^\text{H}_\text{E} (\vec{\sigma}_+ + \vec{\sigma}_-)/\sqrt{2}$ and $\hat{\xi}^\text{V}_\text{E}\vec{V}=i\hat{\xi}^\text{V}_\text{E} (\vec{\sigma}_+ - \vec{\sigma}_-)/\sqrt{2}$ due to noise entering the cavity via the external environment through the $\kappa_\text{ex}$ channel, and $\hat{\xi}^\text{H}_\text{I}\vec{H}=\hat{\xi}^\text{H}_\text{I} (\vec{\sigma}_+ + \vec{\sigma}_-)/\sqrt{2}$ and $\hat{\xi}^\text{V}_\text{I}\vec{V}=i\hat{\xi}^\text{V}_\text{I} (\vec{\sigma}_+ - \vec{\sigma}_-)/\sqrt{2}$ due to internal loss channels within the cavities. We define nominal transmissions $t_\pm = 1+ r_\pm$ for use below. 

The quantum Langevin equations for the cavity modes $\hat{a}^\pm_\text{in}$ and spin operators $\hat{c}_\pm$ now take the form
\begin{subequations}\label{eq:LgvA}
 \begin{align}
  \dot{\hat{a}}_\pm = & (-i\omega_r-\kappa) \hat{a}_\pm -iG\hat{c}_\pm  +\sqrt{\kappa_\text{ex}} \hat{a}^\text{H}_\text{in}\\ \nonumber
  & + \sqrt{\kappa_\text{ex}} \hat{\xi}^\text{H}_\text{E} + \sqrt{\kappa_\text{i}} \hat{\xi}^\text{H}_\text{I}  \;\\ \nonumber
 & \pm i \sqrt{\kappa_\text{ex}} \hat{\xi}^\text{V}_\text{E} \pm i \sqrt{\kappa_\text{i}} \hat{\xi}^\text{V}_\text{I} \;,\\
  \dot{\hat{c}}_\pm = & [-i(D\pm A \pm \delta)-\gamma/2] \hat{c}_\pm -iG\hat{a}_\pm \;,
 \end{align}
\end{subequations} 
where $\kappa = \kappa_\text{ex} + \kappa_\text{i}$ is the total decay rate of the cavity.
 The modes $\hat{c}_\pm$ are Holstein-Primakoff transform of the collective spin operators, and we don't consider noise entering them.
 The internal noise at $\omega = \omega_\text{in}$, $\hat{\xi}^\text{H}_\text{I}$ and $\hat{\xi}^\text{V}_\text{I}$, also enter the output ports through the scattering matrix
\begin{equation}\label{eq:St}
\begin{split}
S_t & = \sqrt{\frac{\kappa_\text{i}}{\kappa_\text{ex}}}
 \begin{pmatrix}
   \frac{t_+ + t_-}{2} & -i \frac{t_+ - t_-}{2} \\
   i\frac{t_+ - t_-}{2} &  \frac{t_+ + t_-}{2}
 \end{pmatrix} \\
& = \sqrt{\frac{\kappa_\text{i}}{\kappa_\text{ex}}}
\begin{pmatrix}
   1+\frac{r_+ + r_-}{2} & i \frac{r_+ - r_-}{2} \\
   -i\frac{r_+ - r_-}{2} &  1+\frac{r_+ + r_-}{2}
 \end{pmatrix}
\;,
\end{split}
\end{equation} 
so that 
\begin{equation} \label{eq:totOut}
 \begin{pmatrix}
  \hat{a}^\text{H}_\text{out} \vec{H} \\
  \hat{a}^\text{V}_\text{out} \vec{V}
 \end{pmatrix}
=S_t
 \begin{pmatrix}
  \hat{\xi}^\text{H}_\text{I} \vec{H} \\
  \hat{\xi}^\text{V}_\text{I} \vec{V}
 \end{pmatrix} 
+ S_r \begin{pmatrix}
  \hat{a}^\text{H}_\text{E} \vec{H} \\
  \hat{a}^\text{V}_\text{E} \vec{V}
 \end{pmatrix}
+ S_r \begin{pmatrix}
  \hat{\xi}^\text{H}_\text{E} \vec{H} \\
  \hat{\xi}^\text{V}_\text{E} \vec{V}
 \end{pmatrix} \;.
\end{equation}
 The output fields are 
\begin{subequations}\nonumber
\begin{align}
 \hat{a}^\pm_\text{out}  & = \frac{r_{\pm}}{\sqrt{2}} (\hat{a}^\text{H}_\text{in} + \hat{\xi}^\text{H}_\text{E} \pm i \hat{\xi}^\text{V}_\text{E} ) \;\\
 & \quad + \sqrt{\frac{\kappa_\text{i}}{\kappa_\text{ex}}} \frac{t_{\pm}}{\sqrt{2}}(\hat{\xi}^\text{H}_\text{I} \pm i \hat{\xi}^\text{V}_\text{I}) \;,\\ \nonumber
\hat{a}^\text{H}_\text{out}  & = \frac{\hat{a}^+_\text{out} + \hat{a}^-_\text{out}}{\sqrt{2}} \;, \\\nonumber
\hat{a}^\text{V}_\text{out}  & = -i \frac{\hat{a}^+_\text{out} - \hat{a}^-_\text{out}}{\sqrt{2}}\;.\nonumber
\end{align}
\end{subequations}
The vertical polarized output includes four contributions: (I) the input probe field, $\hat{a}^\text{H}_\text{in}$; (II) the external H-polarized noise $\hat{\xi}^\text{H}_\text{E}$; (III) the internal H-polarized noise $\hat{\xi}^\text{H}_\text{I}$; (IV) the external V-polarized noise $\hat{\xi}^\text{V}_\text{E}$; (V) the internal V-polarized noise $\hat{\xi}^\text{H}_\text{I}$. The total V-polarized output field is given by
\begin{equation} \label{eq:aoutVA}
\begin{split}
 \hat{a}^\text{V}_\text{out} & = - i r_{\vec{V}\vec{H}} (\hat{a}^\text{H}_\text{in} + \hat{\xi}^\text{H}_\text{E} + \sqrt{\frac{\kappa_\text{i}}{\kappa_\text{ex}}}  \hat{\xi}^\text{H}_\text{I}) \;\\
 & + [r_{\vec{H}\vec{H}} \hat{\xi}^\text{V}_\text{E} + (1+r_{\vec{H}\vec{H}})\sqrt{\frac{\kappa_\text{i}}{\kappa_\text{ex}}} \hat{\xi}^\text{V}_\text{I}]\;.
\end{split}
\end{equation}
Note that $(r_\text{+} - r_\text{-})/2 = (t_\text{+} - t_\text{-})/2$ and $(r_\text{+} + r_\text{-})/2 = -1+ (t_\text{+} + t_\text{-})/2$. 

The whole device is put in an environment with a uniform temperature $T$. The cavity thermal noise operators are defined as $\hat{n}^\text{x}_\text{E} =  \hat{\xi}^{\text{x}\dag}_\text{E} \hat{\xi}^\text{x}_\text{E} $ and $\hat{n}^\text{x}_\text{I} = \hat{\xi}^{\text{x}\dag}_\text{I} \hat{\xi}^\text{x}_\text{I}$ with $\text{x}\in \{H,V\}$. Their quantum expectations are $\langle \hat{n}^\text{x}_\text{E} \rangle  = 2\kappa_\text{ex} \bar{n}_\text{th}$ and $\langle \hat{n}^\text{x}_\text{I} \rangle =  2\kappa_\text{i} \bar{n}_\text{th}$ \cite{PhotonVar}, respectively, with $\bar{n}_\text{th} = 1/(e^{\hbar\omega_r/K_BT}-1)$, $K_B$ is the Boltzman constant. Note that only near-resonant noise enters the cavity and detector. Their variances are $\Delta \hat{n}^{\text{x}2}_\text{E} = (2\kappa_\text{ex})^2 \left(\bar{n}^2_\text{th} + \bar{n}_\text{th}\right)$ and $\Delta \hat{n}^{\text{x}2}_\text{I} = (2\kappa_\text{i})^2 \left(\bar{n}^2_\text{th} + \bar{n}_\text{th}\right)$ \cite{PhotonVar}. We also have $\langle \hat{n}_\text{in}\rangle = 2\kappa_\text{ex}\bar{n}_\text{in}$ and $\Delta \hat{n}^2_\text{in} = (2\kappa_\text{ex})^2 \bar{n}_\text{in}$ for the coherent input field. The quantum expectations and the variances of the $\text{H}$- and $\text{V}$-polarized thermal fields are equal.

 The small magnetic field $\delta B$ to be measured is contained in the measurement operator $\hat{M}= \hat{a}^{\dag\text{V}}_\text{out}\hat{a}^\text{V}_\text{out}$.
 The $\text{V}$-polarized output field is a noisy coherent state with the total thermal noise $\hat{\xi}_\text{tot} = - i r_{\vec{V}\vec{H}} (\hat{\xi}^\text{H}_\text{E} + \sqrt{\frac{\kappa_\text{i}}{\kappa_\text{ex}}}  \hat{\xi}^\text{H}_\text{I}) + [r_{\vec{H}\vec{H}} \hat{\xi}^\text{V}_\text{E} + (1+r_{\vec{H}\vec{H}})\sqrt{\frac{\kappa_\text{i}}{\kappa_\text{ex}}} \hat{\xi}^\text{V}_\text{I}]$ on top of a coherent state $|- i r_{\vec{V}\vec{H}} \langle \hat{a}^\text{H}_\text{in}\rangle\rangle$. We define $\bar{n}_\xi = \langle \hat{\xi}^\dag_\text{tot}\hat{\xi}_\text{tot} \rangle$ and have \cite{PhotonVar}
\begin{subequations}\nonumber
 \begin{align}
 & \bar{n}_\xi  = 2\kappa_\text{ex} P_{\vec{V}}\bar{n}_\text{th} \left[ 1 + \left(\frac{\kappa_\text{i}}{\kappa_\text{ex}}\right)^2 \right] \\ \nonumber 
  & \quad \quad+ 2\kappa_\text{ex}  P_{\vec{H}} \bar{n}_\text{th}\left[1+ \frac{1+P_{\vec{H}} + 2\Re[r_{\vec{H}\vec{H}}]}{P_{\vec{H}}}\left(\frac{\kappa_\text{i}}{\kappa_\text{ex}}\right)^2 \right] \;\\
 & \Delta (\hat{\xi}^\dag_\text{tot}\hat{\xi}_\text{tot})^2  = \bar{n}_\xi^2 + \bar{n}_\xi \;,
 \end{align}
\end{subequations}
where $\Re[x]$ gives the real part of a complex number $x$.
 The quantum expectation of the operator $\hat{M}$ reads 
\begin{equation}\label{eq:EPM}
 \langle \hat{M} \rangle =  2\kappa_\text{ex} P_{\vec{V}} \bar{n}_\text{in} +  \bar{n}_\xi\;,
\end{equation}
where $P_{\vec{V}} = |r_{\vec{V}\vec{H}}|^2$.
Thus, the variance of the output can be evaluated as 
\begin{equation}\nonumber
\begin{split}
 \Delta \hat{M}^2 &  = (2 \kappa_\text{ex} P_{\vec{V}})^2 \bar{n}_\text{in} (2\bar{n}_\xi /2\kappa_\text{ex} P_{\vec{V}} + 1) + (\bar{n}_\xi^2 + \bar{n}_\xi)\;.
\end{split}
\end{equation}
When the mean photon number of thermal noise entering the detector is much larger than unity, and the photons in the probe field is much larger than this thermal noise, i.e. $\bar{n}_\text{in} \gg \bar{n}_\xi \gg 1$, to a good approximation, we have
\begin{equation} \label{eq:VarM}
 \Delta \hat{M}^2 = 4\kappa_\text{ex} P_{\vec{V}} \bar{n}_\text{in} \left(\bar{n}_\xi  + \kappa_\text{ex} P_{\vec{V}} \right) \;,
\end{equation}
and also 
\begin{equation} \label{eq:DM}
 \left( \frac{\partial \langle \hat{M}\rangle}{\partial \delta} \right)^2 = 4\kappa_\text{ex}^2 \left( \frac{\partial P_{\vec{V}}}{\partial \delta}  \right)^2 \bar{n}^2_\text{in} \;.
\end{equation}

Substituting Eqs. (\ref{eq:VarM}) and (\ref{eq:DM}) into Eq. (\ref{eq:VarLim}), and using $\delta = \mu_B g_e \delta B$, $2 \kappa_\text{ex} \tau_\text{ex} =1$, we obtain the sensitivity limit for the multiphoton input
\begin{equation} \nonumber
\begin{split}
 \Delta B_\text{MP} \sqrt{\tau_\text{tot}} & \gtrsim \frac{\sqrt{\tau_\text{m}}}{\sqrt{F_\text{I,V}(\delta)}} \frac{\sqrt{\left(2\bar{n}_\xi  + 2\kappa_\text{ex}P_{\vec{V}}  \right)}}{\sqrt{2\kappa_\text{ex}\bar{n}_\text{in}}} \;\\
 & \gtrsim \frac{1}{\sqrt{F_\text{I,V}(\delta)}} \frac{\sqrt{\mathcal{C}_\text{th} K_B T + P_{\vec{V}} \hbar \omega_r }}{\sqrt{P_\text{in}}} \; ,
\end{split}
\end{equation}
where $F_\text{I,V}(\delta) = (\mu_Bg_e)^2 \frac{1}{P_{\vec{V}}} \left( \frac{\partial P_{\vec{V}}}{\partial \delta}  \right)^2$ is the nominal Fisher information of the vertical-polarized output field and $\mathcal{C}_\text{th} = 2 P_{\vec{V}} \left[1+ \left(\frac{\kappa_\text{i}}{\kappa_\text{ex}} \right)^2 \right] +  2 P_{\vec{H}} \left[1+ \frac{1+P_{\vec{H}}+ 2\Re[r_{\vec{H}\vec{H}}]}{P_{\vec{H}}} \left(\frac{\kappa_\text{i}}{\kappa_\text{ex}} \right)^2\right]$. 
Here we applied the relation $\bar{n}_\text{th}\hbar\omega_r \approx K_B T$ for $\bar{n}_\text{th} \gg 1$. In the case of $\kappa_\text{i} \ll \kappa_\text{ex}$, the limit becomes
\begin{equation} \label{eq:MPLim}
 \Delta B_\text{MP} \sqrt{\tau_\text{tot}} \gtrsim \frac{1}{\sqrt{F_\text{I,V}(\delta)}} \frac{\sqrt{2(P_{\vec{V}} +P_{\vec{H}}) K_B T + P_{\vec{V}} \hbar \omega_r }}{\sqrt{P_\text{in}}} 
\end{equation}

The multiphoton limit of Eqs. (\ref{eq:MPLim}) immediately shows us that the sensitivity is proportional to $\sqrt{T/P_\text{in}}$ for $K_BT \gg \hbar\omega_r$, and can be improved by cooling the resonator or increasing the probe power.

It is very interesting to note that the sensitivity limit in the multiphoton measurement Eq. (\ref{eq:MPLim}), is inversely proportional to $\sqrt{P_\text{in}} \propto \sqrt{\tau_m/\bar{n}_\text{in}}$. It means that the sensitivity cannot be improved by increasing the duration of the probe pulse when keeping the input power constant.

The multiphoton limit also shows an important advantage for sensing using a mw cavity with an input mw field of frequency $\omega_\text{mw}$ over an optical system with $\omega_\text{o}\gg \omega_\text{mw}$. Given the same Fisher information, the input power $P_\text{in}$ and the environmental temperature we have $\hbar \omega_\text{mw}/2\ll K_BT \ll \hbar \omega_\text{o}/2$, the sensitivity can be improved by $\sim\sqrt{\frac{\hbar \omega_\text{o}}{2K_BT}}$ using the mw system in comparison with the optical system. For typical parameters $T=70 ~\text{\kelvin}$ and $\omega_\text{o} = 1.78\times 10^{15} ~\text{\radianpersecond}$ corresponding to the wavelength $\lambda=1064~\text{\nano\meter}$, the improvement factor is about $10$ and can be up to two orders as $T\rightarrow 0$ for $\omega_\text{mw} =2\pi \times 3~ \text{\giga\hertz}$.

% Note that in the above analysis we neglected magnetic noise from the ensemble of spins due to inhomogeneous broadening in the transition energies. If such noise shifts the transition energies for both $\sigma_+$ and $\sigma_-$ transitions in the same direction, we can suppress its affect by adjusting the frequency $\omega_\text{in}$ of the probe field; while if it shifts the energy oppositely, its affect can also be suppressed by the static offset $B_0$.

Our system is too complex to provide an analytical form for the Fisher information and in what follows we calculate it numerically.

\section{Numerical Results}

%\subsection{Conditional probabilities for detection}

\begin{figure}
\centering
\includegraphics[width=0.7\linewidth]{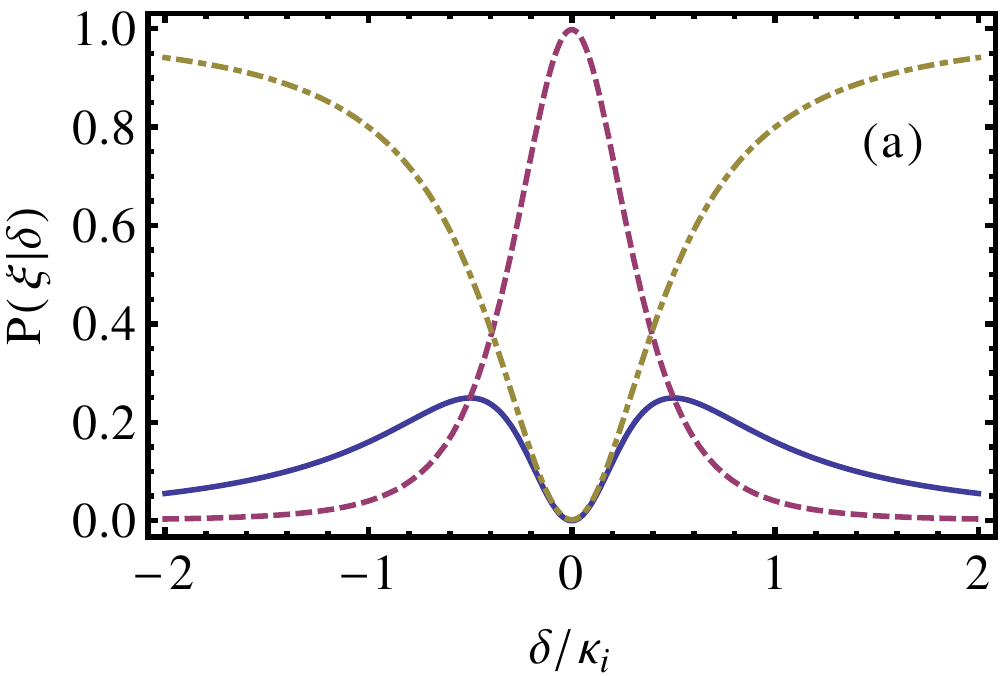} \\
\includegraphics[width=0.7\linewidth]{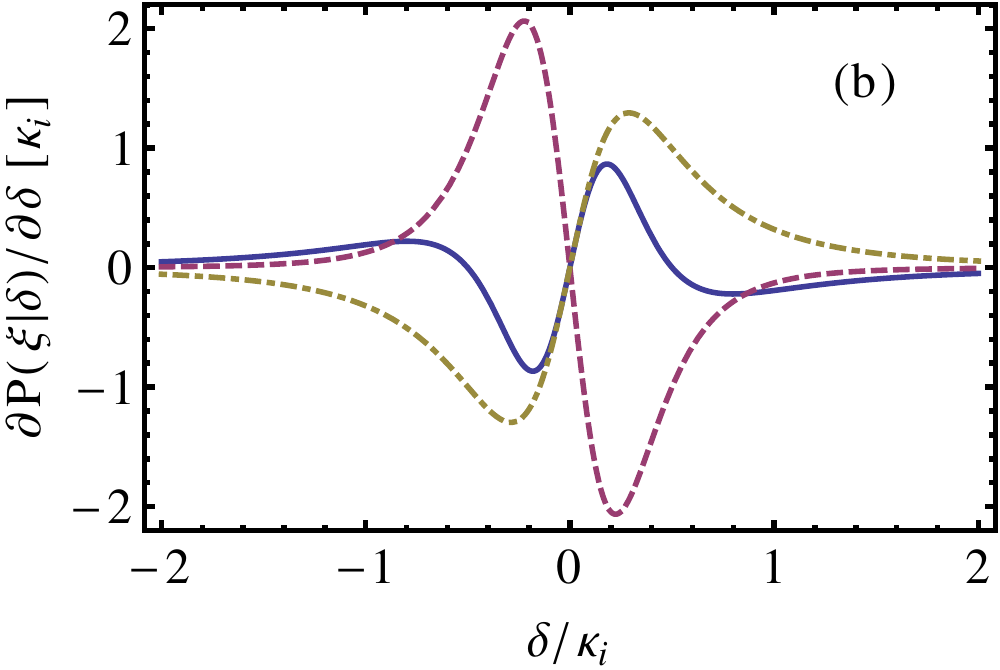} \\
\caption{(Color online) Probabilities $P(\xi|\delta /\kappa_{i})$ (a) and derivation of probabilities $\partial P(\xi|\delta)/\partial \delta$ (b) of three outputs as a function of the magnetic-field induced level shift $\delta$. $\Delta_b=\Delta_a=\Delta_q=0,A=0,\kappa_{ex}=\kappa_{i}, G=\kappa_{i},\gamma=10^{-3}$. Blue lines indicates the probability of the vertical polarization ($P_{\vec{V}}$), red lines for the horizontal polarized outcome ($P_\text{H}$) and green lines for the probability of detecting zero photon  ($P_\varnothing$).
}\label{fig:P}
\end{figure}
Figure \ref{fig:P} shows the probabilities, Eqs. (\ref{eq:Prob}), and their derivatives for the three possible outcomes (horizontal, vertical and no photon). The conditional probabilities $P(\xi|\delta)$ ($\xi=\{\vec{V},\vec{H},\O{}\}$), are found to be symmetric with respect to $\delta=0$. The probability $P(\O{}|\delta)$ has a Lorentzian profile with a dip  $P(\O{}|\delta)=0$ at $\delta=0$. When $\delta \sim 0$, the probability to detect a horizontal polarized photon is nearly unity. When the cavity and the spins are off resonant the probability to detect a vertical polarized photon increases and reaches the maximum $P(\vec{V}|\delta)=0.25$ at $|\delta|=0.494\kappa_\text{i}$, and then decreases again. This nonzero value of $\delta$ indicates that there is an optimal bias with $A=\delta$. Throughout the description below, we will replace the nonzero $\delta$ with the bias $A$. 

The derivatives of these probabilities are antisymmetric with respect to $\delta=0$. They reach  abolsute maxima  at similar positions, $\delta \sim 0.2\kappa_\text{i}$. Therefore, when the spins are biased at $A\sim 0.2\kappa_\text{i}$, our magnetometer is most sensitive to the weak magnetic signal.

%\subsection{Fisher information}
The Fisher information is crucially dependent on the conditional probability derivatives. At $\delta=0$, the Fisher information has a deep drop, while it has a peak of $F_I(\delta/\kappa_{i})=29 \left(\frac{\mu_B g_e}{\kappa_\text{i}}\right)^2$ at $|A|=0.07\kappa_{i}$. The full FWHM defining a bandwidth for measurement is about $0.6\kappa_{i}$. Note that the resolution of a probing system is governed by the product of the average Fisher information and the bandwidth $1/\tau_m$, see Eq. (\ref{eq:SPLimit}). This gives a sensitivity of $\Delta B \sqrt{\tau_\text{total}} \gtrsim 1.2 \times 10^{-11} \sqrt{\kappa_\text{i}}$ for $G=\kappa_\text{ex} = \kappa_\text{i}$ and $\gamma = 10^{-3} \kappa_\text{i}$.

\begin{figure}
\centering
\includegraphics[width=0.7\linewidth]{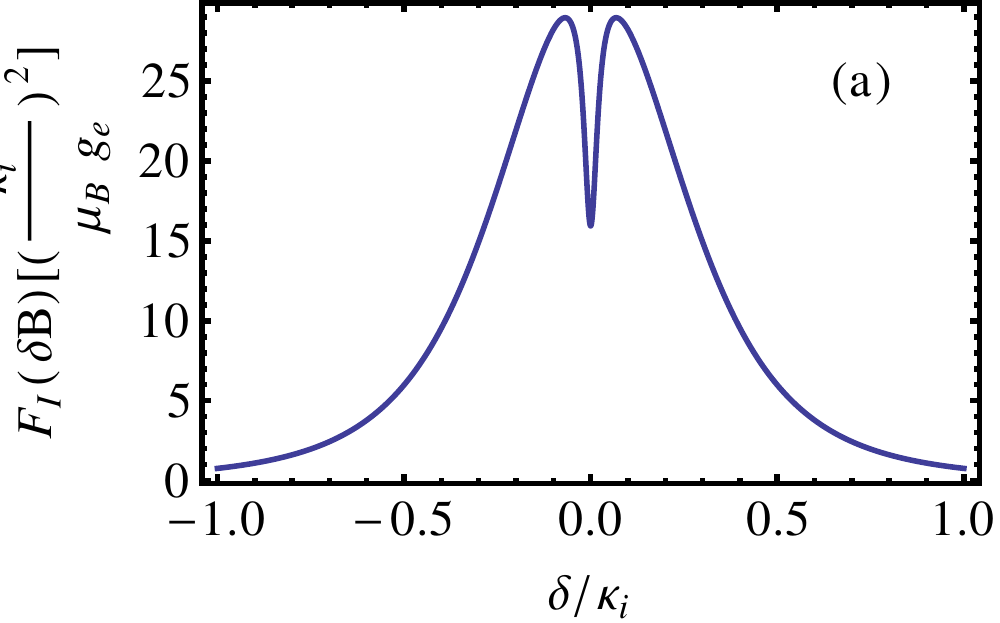}\\
\caption{(Color online) Fisher information as a function of the magnetic-field induced level shift $\delta$. $\Delta_r=\Delta_q=0,A=0,G=\kappa_{ex}=\kappa_\text{i},\gamma=10^{-3}\kappa_\text{i}$. 
}\label{fig:FIdelta}
\end{figure}

%\subsection{Optimal sensitivity limits}

\begin{figure}
\centering
\includegraphics[width=0.7\linewidth]{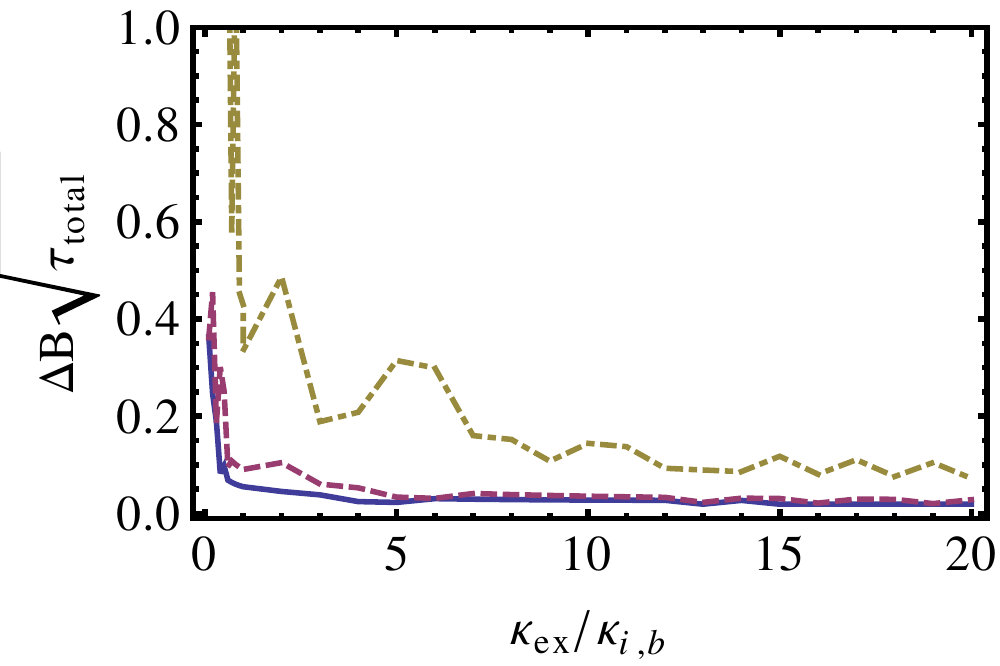}\\
\caption{(Color online) Sensitivity scaled by $\frac{\sqrt{\kappa_\text{i}}}{\mu_Bg_e}$ as a function of the external coupling $\kappa_\text{ex}$ and the cavity-spin coupling $G$. $G/\kappa_\text{i}=(0.02,0.1,0.2)=\text{(solid blue line, dashed red line, dot-dashed yellow line)}$. $\Delta_r=\Delta_q=0$. 
}\label{fig:SPMAR}
\end{figure}

Now we find the optimal sensitivities. As shown in Fig. \ref{fig:SPMAR}, we calculate the sensitivity for the single-photon input as a function of the external coupling $\kappa_\text{ex}$ for different spin-cavity coupling $G$. It can be seen that the optimal sensitivity of $<0.03 \frac{\sqrt{\kappa_\text{i}}}{\mu_Bg_e}$ can be obtained when $G/\kappa_\text{i} \sim 0.1$ and $\kappa_\text{ex}/\kappa_\text{i} >7$.For a large coupling, e.g. $G/\kappa_\text{i}=1$, the sensitivity is low, $>0.1\frac{\sqrt{\kappa_\text{i}}}{\mu_Bg_e}$. A very small coupling, e.g. $G/\kappa_\text{i}=0.02$, the sensitivity becomes lower. For mediate coupling rates, $G/\kappa_\text{i}\sim 0.1$, the sensitivity rapidly decreases to a limit of $0.027\frac{\sqrt{\kappa_\text{i}}}{\mu_Bg_e}$ at $\kappa_\text{ex}/\kappa_\text{i} \approx 10$. 

\begin{figure}
\centering
\includegraphics[width=0.7\linewidth]{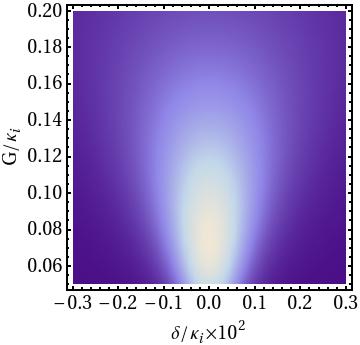}\\
\caption{(Color online) Fisher information scaled by $\left(\frac{\mu_B g_e}{\kappa_\text{i}} \right)^2$ as a function of the spin-cavity coupling $G$ and the magnetic-field induced level shift $\delta$. $\Delta_r=\Delta_q=0$ and $\kappa_\text{ex}=10\kappa_\text{i}$. 
}\label{fig:FIGA}
\end{figure}

To find the optimal spin-cavity coupling $G$, we calculate the Fisher information, shown in Fig. \ref{fig:FIGA}. Clearly, the Fisher information $F_I$ is large when $0.06\leq G/\kappa_\text{i}\leq 0.1$, whereas the width increases from $6.4\times 10^{-4} \kappa_\text{i}$ to $9.6\times 10^{-4} \kappa_\text{i}$. This indicates that the sensitivity is optimal when $G\approx 0.1\kappa_\text{i}$.

\begin{figure}
\centering
\includegraphics[width=0.9\linewidth]{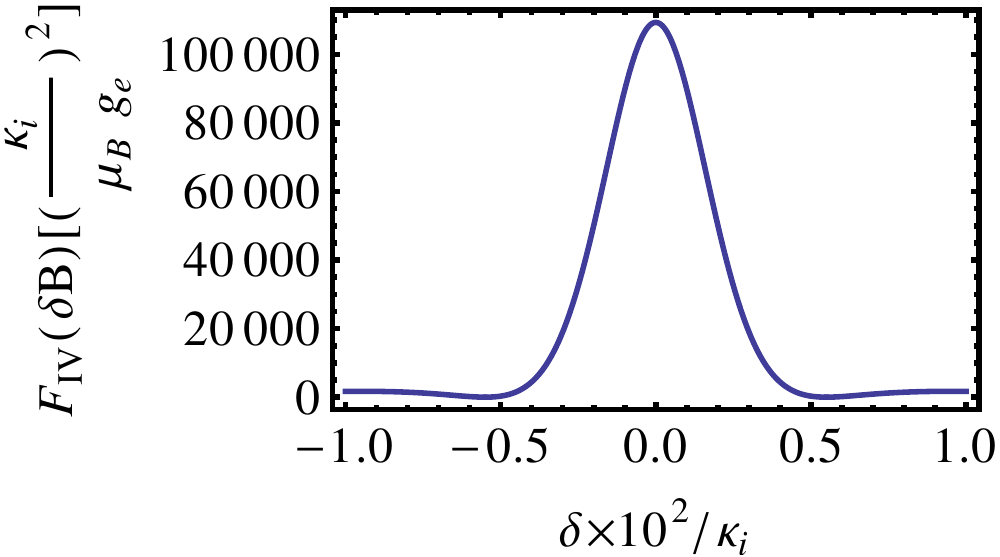}\\
\caption{(Color online) Fisher information as a function of the intercavity coupling $J$ and the detuning $\delta$. $\Delta_b=\Delta_a=\Delta_q=0,A=0,\kappa_\text{ex}/\kappa_\text{i}=10, \gamma/\kappa_\text{i} =10^{-3},G/\kappa_\text{i} =0.1$. 
}\label{fig:MPMFIV}
\end{figure}

For the multiphoton input, we are interested in the maximal Fisher information of the vertical-polarised output. Once known, we can estimate the sensitivity for any environmental temperature and any input power. Figure \ref{fig:MPMFIV} shows the nominal Fisher information of the vertical-polarised output as a function of the detuning $\delta$. The maximum value is about $F_\text{I,V}\approx 10^5 \left(\frac{\mu_B g_e}{\kappa_\text{i}}\right)^2$ yielding $1.92\times 10^{-19} \sqrt{\kappa_\text{i}}$ for $P_\text{in}=1~\nano\watt$ and $T=70~\kelvin$. Note that the FWHM is small, about $4\times 10^{-3} \kappa_\text{i}$. Thus, to achieve the highest sensitivity, the probe pulse need be long.

\section{Discussion of experimental implementation}
The experimental implementation of our magnetometer crucially relies on the realization of a microwave cavity supporting both $\sigma_+$- and $\sigma_-$-polarized (or $\text{H}$- and $\text{V}$-polarized) modes, which are degenerate in frequency. Such microwave cavities require particular designs and have been demonstrated using transmission line resonators \cite{TransLineMWCavity1,TransLineMWCavity2,TransLineMWCavity3} and three-dimensional superconducting Fabry-P\'erot microwave cavities \cite{FPMWcavity1,FPMWcavity2,FPMWcavity3}. The Q factor of the transmission line resonator can be $100-10^5$ ~ \cite{TransLineMWCavity1,TransLineMWCavity4}, while that of the superconducting FP mw cavity can reach $10^{12}$ ~\cite{FPMWcavity1}.  Recently, Zhang et al. developed a photonic crystal(PC) cavity with a small mode volume and a high Q (up to $10^5$), working at microwave frequencies \cite{PCMECavity}. This PC cavity can also possess two degenerate modes with orthogonal polarizations \cite{PCcavity1,PCcavity2,PCcavity3}. In recent works \cite{TransLineMWCavity1,TransLineMWCavity2}, the magnetic transition between the triplet ground states of NV centers has been selectively driven with circularly-polarized microwave fields in microstrip mw resonators. Thus many of the essential components for our scheme have already been experimentally demonstrated.

Now we estimate the optimal sensitivity for detecting a weak magnetic field using an ensemble of NV centers coupled to a microwave cavity. Since the magnetic transition frequency of NV is about $2\pi \times 2.8$ \giga\hertz, we require a microwave cavity with $\omega_r=2\pi \times 2.8$ \giga\hertz. The decay rate $\gamma$ of $|m_s=\pm 1\rangle$ of the ground state triplet has been measured to vary from a few \mega\hertz \cite{NVRT1,NVRT2} to $0.01 \text{\hertz}$ \cite{NVRT3,NVRT4,NVRT5}. Our setup with practical parameters $\gamma=10^{-3}\kappa_\text{i}$, $\kappa_\text{ex} =10 \kappa_\text{i}$ and $G=0.1\kappa_\text{i}$,  using a low-Q
factor cross microstrip resonator with $Q=100$ \cite{TransLineMWCavity1,TransLineMWCavity2,TransLineMWCavity3}, yields $\kappa_\text{i}/2\pi= 28~\mega\hertz$, and can provide a sensitivity of $\Delta B_\text{SP} \sqrt{\tau_\text{total}} \gtrsim 5.2~ \text{\nano\tesla}/\sqrt{\text{\hertz}}$ for a single-photon probe field. If an input power of $P_\text{in}=1~\nano\watt$ is applied and the environmental temperature is fixed to $T=70~\kelvin$, the sensitivity can approach $\Delta B_\text{MP} \sqrt{\tau_\text{total}} \gtrsim 1~ \text{\femto\tesla}/\sqrt{\text{\hertz}}$. Note that sensitivities scales as $\sqrt{\kappa_\text{i}}$. Therefore, if superconducting Fabry-P\'erot cavities with a mediate $Q=10^5$ \cite{FPMWcavity1}, corresponding to $\kappa_\text{i}/2\pi= 28~\kilo\hertz$, is applied, the sensitivities can be improved to $\Delta B_\text{MP} \sqrt{\tau_\text{total}} \gtrsim 32.2~ \text{\atto\tesla}/\sqrt{\text{\hertz}}$ when $P_\text{in}=1~\nano\watt$.

\section{Conclusion}
We have studied the polarization rotation of linear-polarized microwave photons input into microwave cavities coupled to an ensemble of spins. Measuring the Faraday rotation of the output photons provides a method to ultrasensitively measure the strength of magnetic fields. The sensitivity limit of this microwave magnetometer is presented for both single-photon and multiphoton probe fields. The sensitivity of the magnetic field in the multiphoton measurement with an input power of $P_\text{in}=1 ~\nano\watt$ can be tens of $\text{\atto\tesla}/\sqrt{\text{\hertz}}$.

\section*{Acknowledgement}
This work was supported by the Australian Research Council Centre of Excellence in Engineered Quantum Systems (Project CE110001013), and the Macquarie University Research Centre for Quantum Science and Technology.

% \bibliographystyle{apsrev4-1}
% \bibliography{refs}

\end{document}